# Yrast state band of even $^{120-130}$Te isotopes under the framework of interacting boson model -1

I Hossain[1,4*], Hewa Y Abdullah[2], I M Ahmed[3], J Islam[5]

[1]Department of Physics, Rabigh college of Science and Arts, King Abdulaziz University, Rabigh 21911, Saudi Arabia

[2]Department of Physics, College of Science Education, Salahaddin University, Erbil, Krg, Iraq

[3]Department of Physics, College of Education, Mosul University, Mosul, Iraq

[4]Department of Physics, Shahjalal university of Science and Technology, Sylhet, Bangladesh

[5]Department of Physics, Mawlana Bhashani Science and Technology University, Tangail, Bangladesh

*Corresponding author: hossain196977@yahoo.com

Tel: +966-558141913

**Abstract**

In this paper, nuclear structure of yrast state bands of even-even $^{120-130}$Te isotopes has been studied under the frame work of interacting boson model (IBM-1). The theoretical energy levels for Z=52, even N = 68-78 up to $12^+$ state have been obtained by using the MatLab 6.5 computer program. The ratio of the excitation energies of first $4^+$ and first $2^+$ excited states, $R_{4/2}$, has been also calculated for those nuclei. The values of the parameters in the IBM-1 hamiltonian yield the best fit to the experimental energy spectrum. The results are compared with the most recent experimental values where an acceptable degree of agreement is achieved. Moreover, as a measure to quantify the evolution, we have studied the transition rate R = (E2: $L^+ \rightarrow (L-2)^+$ ) / (E2: $2^+ \rightarrow 0^+$) of some of the low-lying quadrupole collective states in comparison to the available experimental data. In this paper, the properties of even $^{120-130}$Te isotopes have been considered to the U(5) transitional region of IBM-1.

Keywords: Nuclear structure, Te isotopes, N= 68-78, U(5), Energy level

*Corresponding author, E-mail: hossain196977@yahoo.com

## 1. Introduction

The interacting boson model-I (IBM-I) is a valuably interactive model developed by Iachello and Arima.[1,2] It has been successful in describing collective nuclear structure by prediction of low-lying states and description of electromagnetic transition rates in the medium mass nuclei. IBM defines six-dimensional space described by in terms of unitary group, U(6). Different reductions of U(6) gives three dynamical symmetry limits known as harmonic oscillator, deformed rotator and asymmetric deformed rotor which are labeled by U(5), SU(3) and O(6) respectively.[3,4]

Even-even tellurium isotopes are part of an interesting region beyond the closed proton shell at Z= 50, while the number of neutrons in the open shell is much larger, which have been commonly considered to exhibit vibration-like properties.[5] Yrast states up to $I^{\pi}= 12^+$ in Z= 52 isotones were found by $\pi h_{11/2}^{+2}$ configurations for the Z=50 closed shell. It is known that low-lying collective quadrupole E2 excitations occur in even-even nuclei Z=52 and N=68-78, which have been studied both theoretically and experimentally. Reorientation effect measurements in $^{120-130}$Te were investigated by Bechara et al.[6] and Barrette et al.,[7] Energy levels, electric quadrupole moments of $^{122}$Te isotopes have been studied within the framework of the semi microscopic model,[8] the two-proton core coupling model[9] and dynamic deformation model.[10]

There are a number of theoretical works discussing intruder configuration and configuration mixing by means of IBM-I around the shell closure Z=50. For instance, empirical spectroscopic study within the configuration mixing calculation in IBM,[11,12] IBM configuration mixing model in strong connection with shell model,[13,14] conventional collective Hamiltonian approach[15,16] and one starting from self-consistent mean-field calculation with microscopic energy density functional.[17] Recently we have been studied the evolution properties of the yrast states for even-even $^{100-110}$Pd isotopes.[18] The electromagnetic reduced transition probabilities and ground-state energy band of even-even $^{104-122}$Cd isotopes were studied.[19,20] The analytic IBM-I calculation of B(E2) values of even-even $^{102-106}$Pd have been confirmed U(5) character.[21]

A basic property of a nucleus is the probability of electric quadrupole (E2) transitions between its low-lying states. In even-even nuclei, the reduced E2 probability *B(E2; $0^+_1$ →$2_1^+$)* from $0^+$ ground state to the first-excited $2^+$ state is especially important, and for a deformed nucleus this probability (denoted here by *B(E2↑)*) depends on the magnitude of the intrinsic quadrupole moment (quadrupole moment of the intrinsic state of the nucleus) and, hence, on deformation.[22] No detailed work has been done on the structure of Te isotopes and it is necessary to carry out calculations that are comparable to experimental results. Our aim in this study is to investigate $^{120-130}$Te isotopes in SU(5) limit and calculate the energy levels of the yrast state band up to $12^+$ level through E2 transition strengths.

**2. Theoretical calculation**

2.1 Ground-state energy levels

The Hamiltonian of the interacting bosons in IBM-1.[23]

$$H = \sum_{i=1}^{N} \varepsilon_i + \sum_{i\langle j}^{N} V_{ij} \qquad (1)$$

where ε is the intrinsic boson energy and *Vij* is the interaction between bosons i and j. In the multi pole form the Hamiltonian.[24]

$$H = \varepsilon \hat{n}_d + a_0 \hat{P}.\hat{P} + a_1 \hat{L}.\hat{L} + a_2 \hat{Q}.\hat{Q} + a_3 \hat{T}_3.\hat{T}_3 + a_4 \hat{T}_4.\hat{T}_4 \qquad (2)$$

Here, $a_0, a_1, a_2, a_3$ and $a_4$ are strength of pairing, angular momentum and multi pole terms. The Hamiltonian as given in Eq.(1) tends to reduces to three limits, the vibration U(5), γ-soft O(6) and the rotational SU(3) nuclei, starting with the unitary group U(6) and finishing with group O(2).[24] In U(5) limit, the effective parameter is ε, in the γ-soft limit, O(6), the effective parameter is the pairing $a_0$, and in the SU(3) limit, the effective parameter is the quadrupole $a_2$. The eigen values for the three limits.[24]

$$U(5): \quad E(n_d, v, L) = \varepsilon n_d + K_1 n_d (n_d + 4) + K_4 v(v+3) + K_5 L(L+1) \qquad (3)$$

$$O(6): E(\sigma, \tau, L) = K_3[N(N+4) - \sigma(\sigma+4)] + K_4 \tau(\tau+3) + K_5 L(L+1) \qquad (4)$$

$$SU(3): \ E(\lambda, \mu, L) = K_2(\lambda^2 + \mu^2 + 3(\lambda + \mu) + \lambda\mu) + K_5 L(L+1) \qquad (5)$$

Here, *K1, K2, K3, K4 and K5* are other forms of strength parameters. Many nuclei have a transition property between two or three of the above limits and their eigen values for the yrast-line.[23]

### 3. Results and discussion

The $^{120-130}$Te isotopes having atomic number Z= 52 and neutron number N = 68-78. A boson number represents the pair of valence nucleons and boson number is counted as the number of collective pairs of valence nucleons. A simple correlation exists between the nuclei showing identical spectra and their valence proton number ($N_p$) and neutron number ($N_n$). The number of valance proton $N_p$ and neutron $N_n$ has a total $N = (N_p + N_n)/2 = n_\pi + n_\nu$ bosons. $^{132}$Sn doubly-magic nucleus is taken as an inert core to find boson number. Boson numbers and the calculated parameters of different levels for even $^{120-130}$Te nucleus in IBM-I are presented in Table 1. All parameters are given in units of keV.

A computing program was written by using the Matlab 6.5 environmental to calculate the energies of some states of the ground state band by applying the eigen value expression for the vibrational nuclei. The energy of ground states band (i.e. $0^+$, $2^+$, $4^+$, $6^+$, $8^+$, $10^+$ and $12^+$) for doubly even isotopes $^{120-130}$Te have been calculated using Eq. (3) by $IBM-1$ calculation. For the ground-state bands only the levels up to spin $12^+$ were considered in the calculation since above this spin value the yrast bands exhibit a backbend phenomena. A suitable free parameters have been determined to find the close excitation-energy of all positive parity levels ($2^+$, $4^+$, $6^+$, $8^+$, $10^+$ and $12^+$) for which a good indication of the spin value exists.[24] However, the seniority number in the ground state band is given by $\nu = n_d = L/2$, therefore the values of this number for $2_1^+$ and $12_1^+$, as example will be as following: $\nu = n_d = L/2 = 1$ for $2_1^+$ and $\nu = n_d = L/2 = 6$ for $12_1^+$ state. The four unknown parameters $\varepsilon$, K1, K4, and K5 were determined by solving Eqs. 2 and 3 for six energy states. Table 1 shows the values of these parameters that have used to calculate the energy of the yrast- states for the isotopes $Z = 52$ and even $N = 68-78$ under this study. Values of these parameters that have been determined so as to reproduce as closely as possible the excitation energy of all yrast state for which a clear indication of the spin value exists. The energy level fits with IBM-1 are presented in table 2 and they are compared with experimental levels.[25-30]. The agreement between calculated theory and experiment is excellent and reproduced well.

3.1. $R_{4/2}$ classifications

It is known that collective dynamics of energies in even-even nuclei are grouped into classes, within each class the ratio of excitation energies of first $4^+$ and first $2^+$ excited states is: spherical vibrator $U(5)$ has $R_{4/2} = 2.00$, γ-unstable rotor O(6) should have $R_{4/2} = 2.5$ and an axially symmetric rotor SU(3) should have $R_{4/2} = 3.33$. We examined U(5) symmetry as $R_{4/2} = 2.07, 209, 2.07, 2.04, 2.01$ and $1.95$ of $^{120}$Te, $^{122}$Te, $^{124}$Te, $^{126}$Te, $^{128}$Te and $^{130}$Te respectively. Fig. 1 shows ratio $R = E(L_1^+) / E(2_1^+)$ values as a function of even neutron of $^{120-130}$Te isotope by experiment results.[25-30] It is shown that all experimental values are closed to solid line of U(5) limit.

In order to carry out the evolution properties we have compared the ratio $R = $ (E2: $L^+ \rightarrow (L-2)^+$ )/ (E2: $2^+ \rightarrow 0^+$) as a function of even neutron number of $^{120-130}$Te isotopes by IBM-I and previous available experimental values and are shown in Fig.2. The values of experimental and IBM-1 are normalized at $2^+$ and $10^+$ levels. It is shown that results of R values remain constant at $R_{4/2}$ from even neutron N= 68-78. However at high spin states the energy ratio decreases as a function of number of neutron in $^{120-130}$Te isotopes. We have found that the calculated values are in good agreement with the previous available experimental results.[25-30]

## 4. Conclusions

In this work, the ground-state energy band of even-even $^{120-130}$Te isotope have been investigated by IBM-I. The present results were compared with previous experimental values. The calculated excitation energies and the experimental ones are in good agreement. The analytic IBM-I calculation of yrast levels up to $12_1^+$ levels of $^{120-130}$Te isotope have been performed in the U(5) character. The results are extremely useful for compiling nuclear data table.

**Acknowledgments**

This work was funded by the Deanship of Scientific Research (DSR), King Abdulaziz University, Jeddah, under Grant No. (662-016-D1434). Therefore, the authors thankfully acknowledge the technical and financial support of DSR.

Table 1. Boson numbers and the calculated parameters for different levels in IBM-1 for $^{120-130}$Te nucleus.

| Nucl. | N | States | Limits | $\varepsilon(keV)$ | $K_1(keV)$ | $K_4(keV)$ | $K_5(keV)$ |
|---|---|---|---|---|---|---|---|
| $^{120}$Te | 8 | 2-12 | U(5) | 495.018 | 9.414 | -3.879 | 5.642 |
| $^{122}$Te | 7 | 2-12 | *U*(5) | 451.183 | 39.386 | -24.092 | 2.058 |
| $^{124}$Te | 6 | 2-12 | U(5) | 514.511 | 52.332 | -42.200 | -.776 |
| $^{126}$Te | 5 | 2-12 | U(5) | 681.474 | 42.504 | -54.834 | -1.385 |
| $^{128}$Te | 4 | 2-12 | U(5) | 791.670 | 32.972 | -59.301 | 3.982 |
| $^{130}$Te | 3 | 2-12 | U(5) | 1141.5 | -26.6 | -48.1 | 3.9 |

Table 2. The experimental[25] and calculated energies and percentage error for levels.

| Nucl. | $I^\pi$ | $E_{exp.}$ (keV) | $E_{cal.}$ (keV) | Δ% |
|---|---|---|---|---|
| $^{120}$Te | $2^+$ | 560.4 | 560.4 | 0.00 |
| | $4^+$ | 1161.1 | 1177.1 | 1.38 |
| | $6^+$ | 1775.1 | 1849.9 | 4.21 |
| | $8^+$ | 2652.4 | 2579.0 | 2.78 |
| | $10^+$ | 3364.3 | 3364.3 | 0.00 |
| | $12^+$ | 4459.4 | 4205.9 | 5.68 |

Table 3. The experimental[26] and calculated energies and percentage error for $^{122}$Te levels.

| Nucl. | $I^\pi$ | $E_{exp.}$ (keV) | $E_{cal.}$ (keV) | Δ% |
|---|---|---|---|---|
| $^{122}$Te | $2^+$ | 560.4 | 560.09 | 0.00 |
| | $4^+$ | 1161.5 | 1175.2 | 0.51 |
| | $6^+$ | 1776.1 | 1833.4 | 4.69 |
| | $8^+$ | 2652.8 | 2538.6 | 4.91 |
| | $10^+$ | 3290.8 | 3290.8 | 0.00 |
| | $12^+$ | 3978.8 | 4090.5 | 5.68 |

Table 4. The experimental[27] and calculated energies and percentage error for $^{124}$Te levels.

| Nucl. | $I^\pi$ | $E_{\text{exp.}}$ (keV) | $E_{\text{cal.}}$ (keV) | Δ% |
|---|---|---|---|---|
| $^{124}$Te | $2^+$ | 602.7 | 602.7 | 0.00 |
|  | $4^+$ | 1248.6 | 1219.5 | 2.33 |
|  | $6^+$ | 1747.0 | 1850.4 | -5.92 |
|  | $8^+$ | 2664.4 | 2495.3 | 6.35 |
|  | $10^+$ | 3154.3 | 3154.4 | 3.45 |
|  | $12^+$ |  | 3827.5 |  |

Table 5. The experimental[28] and calculated energies and percentage error for $^{126}$Te levels.

| Nucl. | $I^\pi$ | $E_{\text{exp.}}$ (keV) | $E_{\text{cal.}}$ (keV) | Δ% |
|---|---|---|---|---|
| $^{126}$Te | $2^+$ | 666.4 | 666.4 | 0.00 |
|  | $4^+$ | 1361.3 | 1297 | 4.70 |
|  | $6^+$ | 1776.2 | 1891.8 | -6.51 |
|  | $8^+$ | 2765.6 | 2451.0 | 11.38 |
|  | $10^+$ | 2974.4 | 2974.4 | 0.00 |
|  | $12^+$ | 3687.9 | 3462.1 | 6.12 |

Table 6. The experimental[29] and calculated energies and percentage error for $^{128}$Te levels.

| Nucl. | $I^{\pi}$ | $E_{exp.}$ (keV) | $E_{cal.}$ (keV) | Δ% |
|---|---|---|---|---|
| $^{128}$Te | $2^+$ | 743.2 | 743.2 | 0.00 |
| | $4^+$ | 1497.0 | 1465.6 | 0.02 |
| | $6^+$ | 1811.2 | 2167.3 | -0.20 |
| | $8^+$ | 2689.4 | 2848.1 | -5.90 |
| | $10^+$ | 2790.7 | 2790.7 | 0.00 |
| | $12^+$ | 3508.1 | 3297.6 | 6.00 |

Table 7. The experimental[30] and calculated energies and percentage error for $^{130}$Te levels.

| Nucl. | $I^{\pi}$ | $E_{exp.}$ (keV) | $E_{cal.}$ (keV) | Δ% |
|---|---|---|---|---|
| $^{130}$Te | $2^+$ | 839.5 | 839.5 | 0.00 |
| | $4^+$ | 1632.9 | 1560.8 | 4.42 |
| | $6^+$ | 1815.3 | 2163.8 | -19.26 |
| | $8^+$ | 2648.6 | 2648.6 | 0.00 |
| | $10^+$ | | 3015.2 | |
| | $12^+$ | | 3263.5 | |

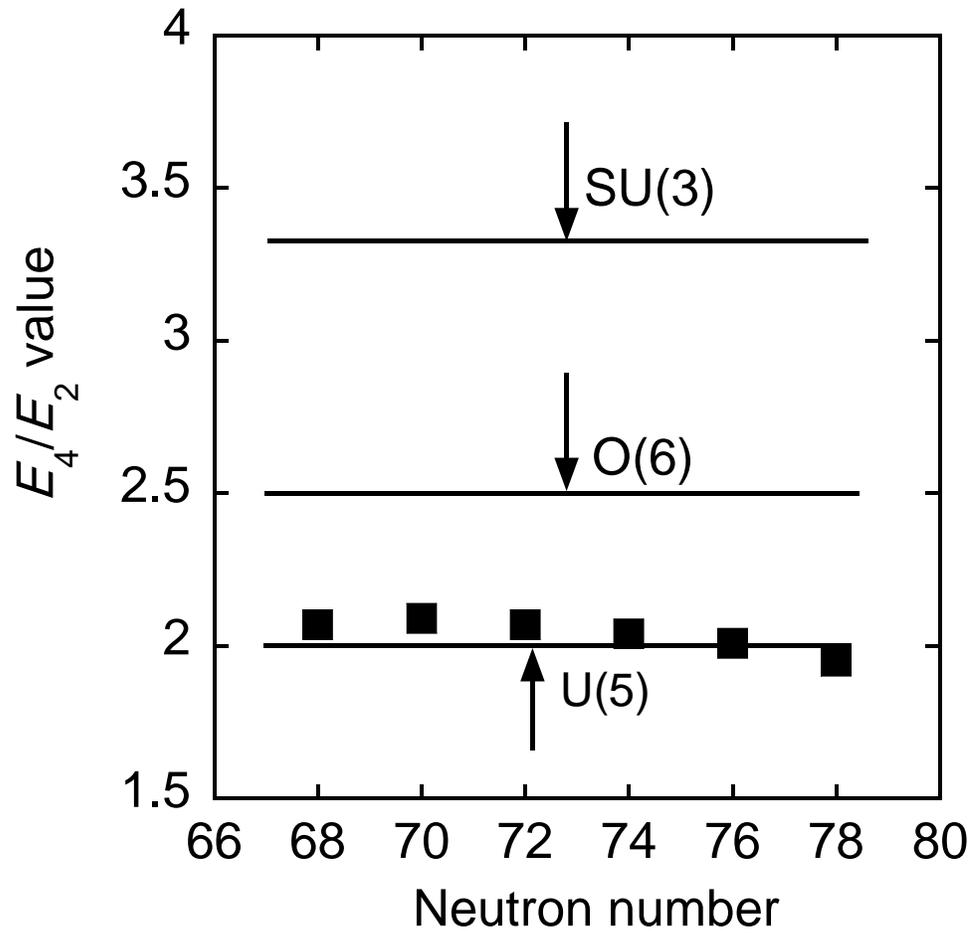

Fig. 1 Plot of ratio R = E($4_1^+$) / E($2_1^+$) experimental values[25-30], U(5), O(6) and SU(3) limit as a function of neutron number of $^{120-130}$Te isotope.

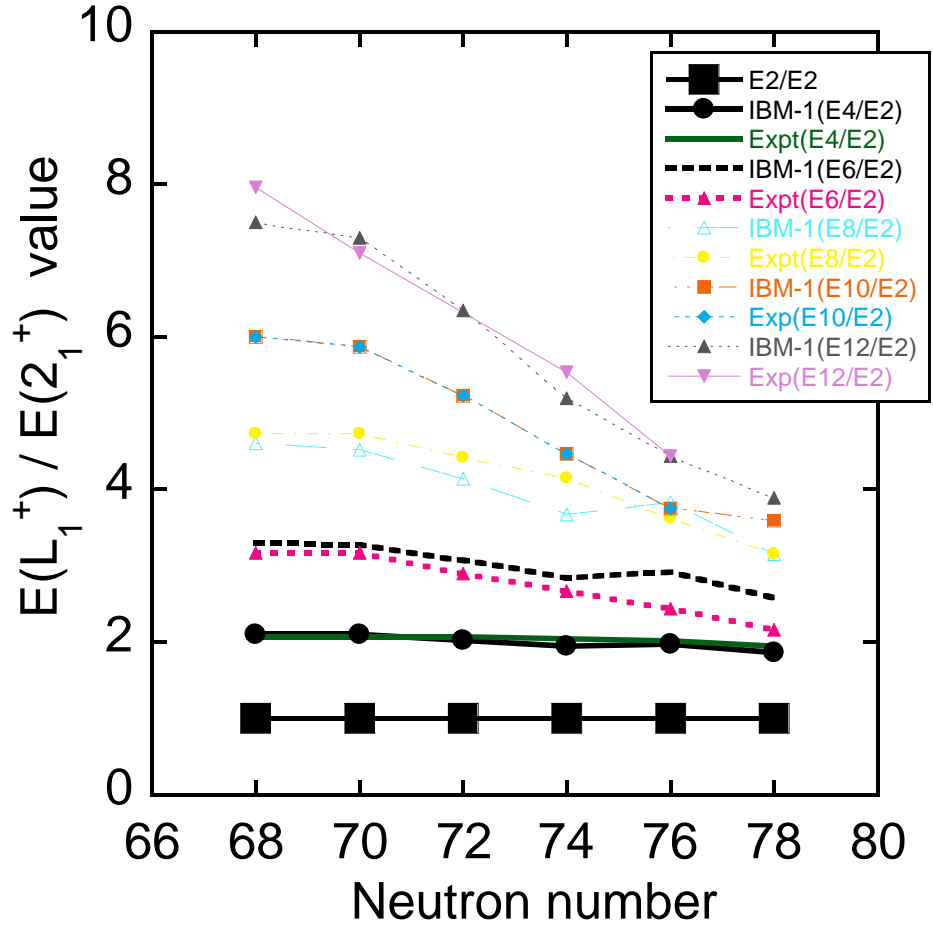

Fig. 2 Plot of ratio R = E($L_1^+$) / E($2_1^+$) values versus even neutron number of $^{120-130}$Te isotope by IBM-I and experiment results.[24] The ratio R = E($L_1^+$) / E($2_1^+$) in ground state bands are normalized to E($2_1^+$).